\documentclass[12pt]{article}
\usepackage{amsmath}
\usepackage{amsfonts}
\usepackage{amssymb}
\usepackage{latexsym}
\usepackage{graphicx}
\usepackage[english]{babel}
\input epsf.sty

\begin{document}
{}~ {}~
\hfill\vbox{\hbox{hep-th/0605048}
}\break

\vskip .6cm \centerline{\Large \bf A Rotating Kaluza-Klein Black
Hole}

\vskip .4cm \centerline{\Large \bf with Squashed Horizons} \vskip
.6cm
\medskip

\vspace*{4.0ex} \centerline{\large \rm Tower Wang} \vspace*{4.0ex}

\centerline{\it Institute of Theoretical Physics, Chinese Academy of
Sciences,}

\centerline{\it P. O. Box 2735£¬ Beijing 100080, China}

\centerline{wangtao218@itp.ac.cn}

\vspace*{5.0ex}

\centerline{\bf Abstract} \bigskip

We find a rotating Kaluza-Klein black hole solution with squashed
$S^3$ horizons in five dimensions. This is a Kerr counterpart of the
charged one found by Ishihara and Matsuno \cite{IM0510} recently.
The space-time is geodesic complete and free of naked singularities.
Its asymptotic structure is a twisted $S^1$ fiber bundle over a
four-dimensional Minkowski space-time. We also study the mass and
thermodynamics of this black hole.

\vfill \eject

\baselineskip=18pt

\tableofcontents

\section{Introduction} \label{s1}

Very recently, Ishihara and Matsuno \cite{IM0510} have found a black
hole solution in the five-dimensional Einstein-Maxwell theory. It is
a charged static black hole with a non-trivial asymptotic topology.
Horizons of the black hole are in the shape of squashed $S^3$, and
the space-time is asymptotically locally flat, approaching a twisted
$S^1$ bundle over a four-dimensional Minkowski space-time. For
simplicity, we will refer to this solution as an ``Ishihara-Matsuno''
black hole or an ``IM''\footnote{Maybe ``RN-IM'' is a more suitable
name to adopt, but anyhow we will not do it here.} black hole for
short, and use the name ``Kerr-IM'' for the Kerr counterpart with
the same asymptotic geometry, etc..

The Ishihara-Matsuno black holes deserve further investigations for
several reasons.

First, string theory tells us that space-time has ten dimensions,
and the idea of large extra dimensions suggests a possibility to
produce black holes smaller than the extra dimensions at colliders
\cite{BF9906,GT0106}. The Ishihara-Matsuno black holes bring us a
new toy model to study. In this space-time, one extra dimension can
be smaller than four uncompactified dimensions, but is much lager
than other dimensions. The size of this large extra dimension is
characterized by an adjustable parameter $r_{\infty}$ \cite{IM0510}.

Second, it is a long-standing problem to define and compute
conserved quantities in a space-time with a non-flat asymptotic
structure or a non-trivial boundary topology. Some breakthroughs
have been made \cite{AD82,AM84,AD9911}, and more attention was
attracted on this problem in recent years
\cite{CLP0510,KLS9906,AR0509,MS0511,HIM0503,CC0601}.
Ishihara-Matsuno black holes have a non-trivial asymptotic
structure, hence provide a new arena to study diverse techniques we
have got. According to calculations performed by Cai et.al.
\cite{CC0603}, the boundary counter-term method \cite{MS0511} and
the generalized Abbott-Deser method \cite{CLP0510} result in the
same mass (energy) for IM black holes, which also satisfies the
first law of thermodynamics. However, as the the compactified
dimension expands, such a mass increases, even in the absence of
five-dimensional black holes. This is a little counterintuitive but
may be explained by comparing IM black holes with Kakuza-Klein
monopoles \cite{Sorkin83}.

Third, from the discussion in \cite{IM0510}, it is clear that
horizons of Ishihara-Matsuno black holes are deformed owing to the
non-trivial asymptotic structure. There is an interesting problem to
ask: What will happen if we switch on angular momenta, the
cosmological constant and Chern-Simons terms \cite{CLP0406} or
replace the black hole by a black ring \cite{ER0110}? We have tried
to answer this question, but only bite a bit, that is, we obtain a
Kerr-IM black hole solution with equal angular momenta in Einstein
theory whose cosmological constant is zero. The Kerr-IM black hole
with non-equal angular momenta is too complicated to struggle with.
When looking for the metric of dS-IM black holes we lose our way. As
far as we know, in five-dimensional asymptotically flat space-time,
charged rotating black holes have not been obtained analytically in
pure Einstein-Maxwell theory yet \cite{KNP0503}. So a Kerr-Newman-IM
black hole solution seems to be equally hard to get. The solution
for Kaluza-Klein multi-black holes see \cite{IKMT0605}.

The organization of this paper is as follows. We produce the Kerr-IM
black hole and identify its asymptotic space-time in the next
section. Then we prove in Section \ref{s3} that such a Kerr-IM
space-time is geodesic complete and free of naked singularities.
Section \ref{s4} and Section \ref{s5} are dedicated to computing the
thermodynamic quantities as well as checking the first law. In
Section \ref{s5}, we also analyze the results and make some comments
on the relation between IM black holes and Kaluza-Klein monopoles. Finally, a
brief summary is given in Section \ref{s6}. Through the entire paper
we exploit the abbreviations introduced at the beginning of this
section.

\section{Kerr-IM Black Holes}\label{s2}
\subsection{A Prescription to Squash Black Holes}\label{ss21}

By observing the Kaluza-Klein black hole with squashed horizons
\cite{IM0510}, and comparing it with the five-dimensional
Reissner-Nordstr\"om black hole, we guess that there is a general
prescription which transforms some known five-dimensional black hole
into a new one with squashed horizons. We further conjecture a
detailed ``squashing transformation'' as the following

\begin{enumerate}
\item Write the original black hole metric in terms of the left-invariant
Maurer-Cartan 1-forms on $S^3$
\begin{eqnarray}\label{Mau-Car}
\nonumber \sigma_1&=&-sin\psi d\theta+cos\psi sin\theta d\phi\\
\nonumber \sigma_2&=&cos\psi d\theta+sin\psi sin\theta d\phi\\
\sigma_3&=&d\psi+cos\theta d\phi
\end{eqnarray}
where
\begin{equation}\label{Euler}
0<\theta <\pi,~~\quad0<\phi<2\pi,~~\quad0<\psi<4\pi
\end{equation}
are Euler angles \cite{CC0603}. The Euler angles are related to
ordinary angles ($\theta_o$, $\phi_o$, $\psi_o$) via \cite{CLP0406}
\begin{equation}\label{Eul ord}
\psi_o-\phi_o=\phi,~~\quad\psi_o+\phi_o=\psi,~~\quad\theta_o={1\over2}\theta
\end{equation}
\item Modify the metric as
\begin{equation}\label{squa transf}
dr\rightarrow kdr,~~\quad\sigma_1\rightarrow \sqrt{k}\sigma_1,~~\quad\sigma_2\rightarrow \sqrt{k}\sigma_2
\end{equation}
while $k$ is a function of $r$ in the form
\begin{equation}\label{squa param}
k(r)={{(r_{\infty}^2-r_{+}^2)(r_{\infty}^2-r_{-}^2)}\over{(r_{\infty}^2-r^2)^2}}
\end{equation}
$r=r_{+}$ corresponds to the outer horizon and $r=r_{-}$ the inner
horizon. $r_{\infty}$ characterizes the size of a $S^1$ fiber at
infinity. We impose to them a constraint $0\leq r_{-}\leq r_{+}<r_{\infty}$.
\end{enumerate}

When we apply this procedure to Kerr black holes with non-equal angular momenta, it
gives a metric too complicated to struggle with. So in the rest
parts of this article, we focus on the squashed Kerr black holes
with equal angular momenta, to which we sometimes refer as
``Kerr-IM'' black holes.

\subsection{From Kerr to Squashed Kerr}\label{ss22}

In terms of Meurer-Cartan 1-forms, a five-dimensional Kerr black
hole with two equal angular momenta takes the form
\begin{equation}\label{kerr metric}
ds^2=-dt^2+{\Sigma\over\Delta}dr^2+{{r^2+a^2}\over4}(\sigma_1^2+\sigma_2^2+\sigma_3^2)+{m\over{r^2+a^2}}(dt-{a\over2}\sigma_3)^2
\end{equation}
The parameters are given by
\begin{eqnarray}\label{kerr param}
\nonumber \Sigma&=&r^2(r^2+a^2)\\
\Delta&=&(r^2+a^2)^2-mr^2
\end{eqnarray}

Taking the second step of the ``squashing transformation'' we have
described in Subsection \ref{ss21}, with
\begin{equation}\label{squa param2}
k(r)={{(r_{\infty}^2+a^2)^2-mr_{\infty}^2}\over{(r_{\infty}^2-r^2)^2}}
\end{equation}
it follows that a Kerr black hole with squashed horizons is
\begin{equation}\label{squa metric1}
ds^2=-dt^2+{\Sigma\over\Delta}k^2dr^2+{{r^2+a^2}\over4}[k(\sigma_1^2+\sigma_2^2)+\sigma_3^2]+{m\over{r^2+a^2}}(dt-{a\over2}\sigma_3)^2
\end{equation}

It is straightforward to check that metric (\ref{squa metric1})
satisfies the vacuum Einstein equation
\begin{equation}\label{Einstein equation}
R_{\mu\nu}-{1\over 2}Rg_{\mu\nu}=0
\end{equation}

We set the order of parameters as $0<r_{-}\leq r_{+}<r_{\infty}$,
with an outer horizon at $r=r_{+}$ and an inner horizon at
$r=r_{-}$, and restrict $r$ within the range $0<r<r_{\infty}$, then
metric (\ref{squa metric1}) is geodesic complete and has no naked
singularity, as we will establish in Section \ref{s3}. In
particular, if we take the limit $r_{\infty}\rightarrow\infty$ and
at the same time concentrate on a region with a finite value of
${r_{\infty}-r}\over{r_{\infty}}$, the Kerr-IM metric (\ref{squa
metric1}) will reduce to a five-dimensional Kerr metric. This is
similar to the IM black hole, which reduces to a five-dimensional
Reissner-Nordstr\"om black hole in the same limit \cite{IM0510}.

\subsection{Asymptotic Structure}\label{ss23}

In order to see the asymptotic structure, we introduce a new radial
coordinate $\rho$ as
\begin{equation}\label{rho}
\rho=\rho_0{r^2\over{r_{\infty}^2-r^2}}
\end{equation}
with
\begin{eqnarray}\label{rho0}
\nonumber \rho_0^2&=&{k_0\over 4}(r_{\infty}^2+a^2)\\
k_0&=&k(r=0)={{(r_{\infty}^2+a^2)^2-mr_{\infty}^2}\over{r_{\infty}^4}}
\end{eqnarray}
and rewrite the metric (\ref{squa metric1}) as
\begin{equation}\label{squa metric2}
ds^2=-dt^2+Ud\rho^2+R^2(\sigma_1^2+\sigma_2^2)+W^2\sigma_3^2+V(dt-{a\over2}\sigma_3)^2
\end{equation}
where $K,V,W,R$ and $U$ are functions of $\rho$ in the form
\begin{eqnarray}\label{asym param}
\nonumber K^2&=&{{\rho+\rho_0}\over{\rho+{{a^2}\over{r_{\infty}^2+a^2}}\rho_0}}\\
\nonumber V&=&{{m}\over{r_{\infty}^2+a^2}}K^2\\
\nonumber W^2&=&{{r_{\infty}^2+a^2}\over{4K^2}}={m\over{4V}}\\
\nonumber R^2&=&{{(\rho+\rho_0)^2}\over{K^2}}\\
U&=&({{r_{\infty}^2}\over{r_{\infty}^2+a^2}})^2\times{{\rho_0^2}\over{W^2-{{r_{\infty}^2}\over{4}}{{\rho}\over{\rho+\rho_0}}V}}
\end{eqnarray}

In the limit $\rho\rightarrow\infty$, i.e., $r\rightarrow
r_{\infty}$, it approaches
\begin{equation}\label{asym metric1}
ds^2=-dt^2+d\rho^2+\rho^2(\sigma_1^2+\sigma_2^2)+{{r_{\infty}^2+a^2}\over4}\sigma_3^2+{m\over{r_{\infty}^2+a^2}}(dt-{a\over2}\sigma_3)^2
\end{equation}

At first sight, the cross-term  between $dt$ and $\sigma_3$ appears
to imply a pathology of the space-time. Nevertheless, if we change
the coordinates as
\begin{eqnarray}\label{psit}
\nonumber \tilde{\psi}&=&\psi-{{2ma}\over{(r_{\infty}^2+a^2)^2+ma^2}}t\\
\tilde{t}&=&\sqrt{{(r_{\infty}^2+a^2)^2-mr_{\infty}^2}\over{(r_{\infty}^2+a^2)^2+ma^2}}~t
\end{eqnarray}
and take the notation $\tilde{\sigma}_3=d\tilde{\psi}+cos\theta
d\phi$, the asymptotic space-time is actually all right
\begin{equation}\label{asym metric2}
ds^2=-d\tilde{t}^2+d\rho^2+\rho^2(\sigma_1^2+\sigma_2^2)+{{(r_{\infty}^2+a^2)^2+ma^2}\over{4(r_{\infty}^2+a^2)}}\tilde{\sigma}_3^2
\end{equation}
This asymptotic topology is the same as that of the IM space-time
\cite{IM0510}: a twisted $S^1$ bundle over a four-dimensional
Minkowski space-time. We have expected it because the asymptotic
structure should not be affected by angular momenta or any other
charges of black holes. From the same point of view, we predict that
the asymptotic geometry ought to be a twisted $S^1$ bundle over a
four-dimensional Anti-de Sitter space-time for a AdS-IM black hole
if it really exists.

At infinity, the size of the compactified dimension is controlled by
$r_{\infty}$, $m$ and $a$ together rather than $r_{\infty}$ alone,
implied by metric (\ref{asym metric2}). One should not be bothered
by this illusion of parametrization. A better parameter may be
obtained if we trade $r_{\infty}^2$ for
\begin{equation}\label{r red}
{r'}_{\infty}^2={{(r_{\infty}^2+a^2)^2+ma^2}\over{r_{\infty}^2+a^2}}
\end{equation}
in which the size of the compactified dimension is apparently
decoupled from $m$ and $a$, but that will only make our formulas
more scattered. In our expressions, for brevity, we will always use
the parameter $r_{\infty}$ instead of ${r'}_{\infty}$. But one
should keep in mind that the geometric interpretation is clearer for
${r'}_{\infty}$ than for $r_{\infty}$.

We stress that the coordinate transformation (\ref{rho}) is valid
only for a finite $r_{\infty}$ value, i.e., $r_{\infty}<\infty$. In
the limit $r_{\infty}\rightarrow\infty$, one merely finds
$\rho\rightarrow0$. Thus we will carry out most computations based
on the metric in the form (\ref{squa metric1}) instead of (\ref{squa
metric2}). But the calculations of mass and angular momenta are
inconvenient from (\ref{squa metric1}), so we will in Section
\ref{s5} resort to the coordinate $\rho$ introduced in (\ref{rho}).
We believe the calculations should always be right for a finite
$r_{\infty}$, given that the boundary counter-term method and the
generalized Abbott-Deser method work.

\section{Absence of Naked Singularities}\label{s3}
\subsection{Ingoing Eddington Coordinates}\label{ss31}

The metric (\ref{squa metric1}) beaks down at $\Delta=0$, i.e.,
$r=r_{\pm}$. These singularities can be removed by a coordinate
transformation
\begin{eqnarray}\label{edd transf}
\nonumber dv&=&dt+{{k(r^2+a^2)^2}\over\Delta}dr\\
d\chi&=&d\psi+{{2ak(r^2+a^2)}\over\Delta}dr
\end{eqnarray}

In the new coordinates ($v$, $r$, $\theta$, $\phi$, $\chi$), metric
(\ref{squa metric1}) turns to
\begin{equation}\label{edd metric}
ds^2=-dv^2+{{r^2+a^2}\over4}[k(\sigma_1^2+\sigma_2^2)+{\sigma_3'}^2]+{m\over{r^2+a^2}}(dv-{a\over2}\sigma_3')^2+2k(dv-{a\over2}\sigma_3')dr
\end{equation}
in which we make use of a notation $\sigma_3'=d\chi+cos\theta$. The
new coordinates are nothing but the ingoing Eddington coordinates.
The metric is regular now at $r=r_{\pm}$.

In fact, $r=r_{\pm}$ are where Killing horizons sit. We will only
discuss outer horizon $r=r_{+}$ in this article. By exchanging
$r_{+}\leftrightarrow r_{-}$ one immediately gets the results for
inner horizon $r=r_{-}$.

From (\ref{edd metric}) we can write down the Killing vector of the
event horizon at $r=r_{+}$
\begin{eqnarray}\label{kill vector1}
\nonumber \xi^{\mu}\partial_{\mu}&=&\kappa_0(\partial_v+\Omega_H\partial_\chi)\\
\Omega_H&=&{{2a}\over{r_{+}^2+a^2}}
\end{eqnarray}
where $\kappa_0$ is a normalization constant to be determined in
Section \ref{s4}. In coordinates ($t$, $r$, $\theta$, $\phi$,
$\psi$), the Killing vector takes the form
\begin{equation}\label{kill vector2}
\xi^{\mu}\partial_{\mu}=\kappa_0(\partial_t+\Omega_H\partial_\psi)
\end{equation}
This result may also be worked out from (\ref{squa metric1})
directly.

\subsection{Geodesic Completeness}\label{ss32}

In the following, we will prove that the space-time of a Kerr-IM
black hole is geodesic complete. The same recipe will be efficacious
for the IM black hole found by H. Ishihara et.al. \cite{IM0510}.

The action of a particle with mass
$m_{part}$ is given by \cite{Weinberg}
\begin{equation}\label{action1}
I=-m_{part}\int_{-\infty}^{+\infty}d\lambda\left[-g_{\mu\nu}(x(\lambda)){{dx^{\mu}(\lambda)}\over{d\lambda}}{{dx^{\nu}(\lambda)}\over{d\lambda}}\right]^{1\over2}
\end{equation}
Inserting (\ref{squa metric1}), and taking $\lambda=t$, it becomes
\begin{eqnarray}\label{action2}
\nonumber I&=&\int_{-\infty}^{+\infty}dtL\\
\nonumber &=&-m_{part}\int_{-\infty}^{+\infty}dt\left[1-{\Sigma\over\Delta}k^2\dot{r}^2-{{r^2+a^2}\over4}k\dot{\theta}^2-{{r^2+a^2}\over4}ksin^2\theta\dot{\phi}^2\right.\\
\nonumber &&\left.-{{r^2+a^2}\over4}(\dot{\psi}+cos\theta\dot{\phi})^2-{m\over{r^2+a^2}}(1-{a\over2}\dot{\psi}-{a\over2}cos\theta\dot{\phi})^2\right]^{1\over2}\\
\nonumber &=&-m_{part}\int_0^{r_{\infty}}{{dr}\over{\dot{r}}}\left[1-{\Sigma\over\Delta}k^2\dot{r}^2-{{r^2+a^2}\over4}ksin^2\theta\dot{\phi}^2\right.\\
&&\left.-{{r^2+a^2}\over4}(\dot{\psi}+cos\theta\dot{\phi})^2-{m\over{r^2+a^2}}(1-{a\over2}\dot{\psi}-{a\over2}cos\theta\dot{\phi})^2\right]^{1\over2}
\end{eqnarray}
in which a dot denotes a derivative with respect to $t$, and we have
assumed $\dot{\theta}=0$. As will be shown later, this assumption is
consistent with geodesic equations.

In order to be brief, we introduce a notation
\begin{eqnarray}\label{v}
\nonumber u&=&\left[-g_{\mu\nu}(x(\lambda)){{dx^{\mu}(\lambda)}\over{d\lambda}}{{dx^{\nu}(\lambda)}\over{d\lambda}}\right]^{1\over2}\\
\nonumber &=&\left[1-{\Sigma\over\Delta}k^2\dot{r}^2-{{r^2+a^2}\over4}ksin^2\theta\dot{\phi}^2\right.\\
&&\left.-{{r^2+a^2}\over4}(\dot{\psi}+cos\theta\dot{\phi})^2-{m\over{r^2+a^2}}(1-{a\over2}\dot{\psi}-{a\over2}cos\theta\dot{\phi})^2\right]^{1\over2}
\end{eqnarray}

Then the Hamiltonian is
\begin{eqnarray}\label{Hamiltonian}
\nonumber H&=&E\\
\nonumber &=&{{\partial L}\over{\partial \dot{x}^{\mu}}}\dot{x}^{\mu}-L\\
\nonumber &=&{{m_{part}}\over u}\left[{\Sigma\over\Delta}k^2\dot{r}^2+{{r^2+a^2}\over4}ksin^2\theta\dot{\phi}^2+{{r^2+a^2}\over4}(\dot{\psi}+cos\theta\dot{\phi})^2\right.\\
\nonumber &&\left.+{m\over{r^2+a^2}}(1-{a\over2}\dot{\psi}-{a\over2}cos\theta\dot{\phi})(-{a\over2}\dot{\psi}-{a\over2}cos\theta\dot{\phi})\right]+m_{part}u\\
&=&{{m_{part}}\over{u}}\left[1-{m\over{r^2+a^2}}(1-{a\over2}\dot{\psi}-{a\over2}cos\theta\dot{\phi})\right]
\end{eqnarray}
For a stationary space-time the energy $E$ is a constant, and the
first integrals of $\phi$ and $\psi$ are found to be
\begin{eqnarray}\label{init integrals}
\nonumber P_{\phi}&=&{{\partial L}\over{\partial \dot{\phi}}}\\
\nonumber &=&{{m_{part}}\over u}\left[{{r^2+a^2}\over4}ksin^2\theta\dot{\phi}+{{r^2+a^2}\over4}cos\theta(\dot{\psi}+cos\theta\dot{\phi})\right.\\
\nonumber &&\left.-{m\over{r^2+a^2}}{a\over2}cos\theta(1-{a\over2}\dot{\psi}-{a\over2}cos\theta\dot{\phi})\right]\\
\nonumber P_{\psi}&=&{{\partial L}\over{\partial \dot{\psi}}}\\
&=&{{m_{part}}\over
u}\left[{{r^2+a^2}\over4}(\dot{\psi}+cos\theta\dot{\phi})-{m\over{r^2+a^2}}{a\over2}(1-{a\over2}\dot{\psi}-{a\over2}cos\theta\dot{\phi})\right]
\end{eqnarray}
We will assume $P_{\phi}=0,P_{\psi}=0$ and show the consistency a
little later. Using these assumptions to solve (\ref{init
integrals}), we get a clean result
\begin{eqnarray}\label{dotphsi}
\nonumber \dot{\phi}&=&0\\
\dot{\psi}&=&{{2ma}\over{(r^2+a^2)^2+ma^2}}
\end{eqnarray}
Combining (\ref{Hamiltonian}) and (\ref{dotphsi}) yields
\begin{eqnarray}
\nonumber u&=&{{m_{part}}\over{E}}{{(r^2+a^2)^2-mr^2}\over{(r^2+a^2)^2+ma^2}}\\
\dot{r}&=&{1\over{k}}{{(r^2+a^2)^2-mr^2}\over{(r^2+a^2)^2+ma^2}}\sqrt{{{(E^2-m_{part}^2)(r^2+a^2)^2+m(m_{part}^2r^2+E^2a^2)}\over{E^2r^2(r^2+a^2)}}}
\end{eqnarray}

At last, the integration of the proper time $\tau$ is
\cite{Weinberg}
\begin{eqnarray}\label{time}
\nonumber \int_{-\infty}^{+\infty}d\tau&=&\int_{-\infty}^{+\infty}d\lambda\left[-g_{\mu\nu}(x(\lambda)){{dx^{\mu}(\lambda)}\over{d\lambda}}{{dx^{\nu}(\lambda)}\over{d\lambda}}\right]^{1\over2}\\
\nonumber &=&\int_0^{r_{\infty}}{{dr}\over{\dot{r}}}u\\
\nonumber &=&{{m_{part}}\over{E}}\int_0^{r_{\infty}}dr{{(r_{\infty}^2+a^2)^2-mr_{\infty}^2}\over{(r_{\infty}^2-r^2)^2}}\\
&&\times\sqrt{{{E^2r^2(r^2+a^2)}\over{(E^2-m_{part}^2)(r^2+a^2)^2+m(m_{part}^2r^2+E^2a^2)}}}
\end{eqnarray}
For $E^2>m_{part}^2$ this integration is divergent when and only
when $r$ approaches $r_{\infty}$, that is, for a particle with
enough energy, their geodesics can be extended to all values of
$\tau$. Accordingly we can expect the Kerr-IM space-time is geodesic
complete, and it is clear now that $r\rightarrow r_{\infty}$ indeed
corresponds to spatial infinity.

In the above, we have assumed $\dot{\theta}=0$, $P_{\phi}=0$ and
$P_{\psi}=0$. To be rigorous, we now demonstrate that these
assumptions are consistent with geodesic equations. Christoffel
connections are easy to derive from metric (\ref{squa metric1}), and
we find the expected vanishing components
\begin{equation}\label{connection}
\Gamma^{\theta}_{\mu\nu}=\Gamma^{\phi}_{\mu\nu}=0~~\quad(\mu,\nu=t,r,\psi)
\end{equation}
This is consistent with our earlier assumptions $\dot{\theta}=0$ ,
$P_{\phi}=0$, $P_{\psi}=0$. Note that $P_{\phi}-P_{\psi}cos\theta=0$
is equivalent to $\dot{\phi}=0$.

\section{Some Thermodynamic Quantities}\label{s4}
\subsection{Temperature}\label{ss41}

To discuss the surface gravity and the temperature, it is convenient
to rewrite metric (\ref{squa metric1}) by defining coordinates
$\tilde{\psi}$ and $\tilde{t}$ exactly as (\ref{psit}), then the
Kerr-IM metric (\ref{squa metric1}) and Killing vector (\ref{kill
vector2}) have the form
\begin{eqnarray}\label{squa metric3}
\nonumber ds^2&=&-\tilde{g}_{00}d\tilde{t}^2+{\Sigma\over\Delta}k^2dr^2+{{r^2+a^2}\over4}k(\sigma_1^2+\sigma_2^2)+{{(r^2+a^2)^2+ma^2}\over4(r^2+a^2)}(\tilde{\sigma}_3-\tilde{\omega}d\tilde{t})^2\\
\xi^{\mu}\partial_{\mu}&=&\tilde{\kappa}_0(\partial_{\tilde{t}}+\tilde{\Omega}_H\partial_{\tilde{\psi}})
\end{eqnarray}
and
\begin{eqnarray}\label{g00}
\nonumber \tilde{g}_{00}&=&{{(r^2+a^2)^2-mr^2}\over{(r^2+a^2)^2+ma^2}}\times{{(r_{\infty}^2+a^2)^2+ma^2}\over{(r_{\infty}^2+a^2)^2-mr_{\infty}^2}}\\
\nonumber \tilde{\omega}&=&\left[{{2ma}\over{(r^2+a^2)^2+ma^2}}-{{2ma}\over{(r_{\infty}^2+a^2)^2+ma^2}}\right]\times\sqrt{{{(r_{\infty}^2+a^2)^2+ma^2}\over{(r_{\infty}^2+a^2)^2-mr_{\infty}^2}}}\\
\nonumber \tilde{\kappa}_0&=&\kappa_0\sqrt{{(r_{\infty}^2+a^2)^2-mr_{\infty}^2}\over{(r_{\infty}^2+a^2)^2+ma^2}}\\
\tilde{\Omega}_H&=&\left[\Omega_H-{{2ma}\over{(r_{\infty}^2+a^2)^2+ma^2}}\right]\times\sqrt{{{(r_{\infty}^2+a^2)^2+ma^2}\over{(r_{\infty}^2+a^2)^2-mr_{\infty}^2}}}
\end{eqnarray}
This form of metric has the merit that $\tilde{g}_{00}\rightarrow1$
and $\tilde{\omega}\rightarrow0$ at spatial infinity $r\rightarrow
r_{\infty}$.

Correspondingly the two-dimensional Euclidean Rindler space-time is
\begin{equation}\label{rind metric}
ds_E^2={\Sigma\over\Delta}k^2dr^2+\tilde{g}_{00}d\tilde{\tau}^2
\end{equation}
It has a conical singularity at $r=r_{+}$ until we make a periodic
identification $\tilde{\tau}\sim\tilde{\tau}+{{2\pi}\over{\kappa}}$,
with
\begin{eqnarray}\label{surf grav}
\nonumber \kappa&=&{{r_{+}-r_{-}}\over{r_{+}^2+a^2}}\times{1\over{k(r_{+})}}\times\sqrt{{{(r_{\infty}^2+a^2)^2+ma^2}\over{(r_{\infty}^2+a^2)^2-mr_{\infty}^2}}}\\
&=&{{(r_{+}-r_{-})\sqrt{(r_{\infty}^2+r_{+}r_{-})^2+r_{+}r_{-}(r_{+}+r_{-})^2}}\over{r_{+}(r_{+}+r_{-})(r_{\infty}^2-r_{-}^2)}}\times\sqrt{{r_{\infty}^2-r_{+}^2}\over{r_{\infty}^2-r_{-}^2}}
\end{eqnarray}

The same result can also be derived from (\ref{squa metric1}) and
(\ref{kill vector2}), using
$\xi\cdot\nabla\xi^{\mu}\mid_{r=r_{\pm}}=\kappa_{\pm}\xi^{\mu}$, and
choosing a normalization to ensure
$\tilde{g}_{00}\xi^{\tilde{t}}\xi^{\tilde{t}}\rightarrow1$ as
$r\rightarrow r_{\infty}$, that is
\begin{equation}\label{norm}
\kappa_0=\sqrt{{{(r_{\infty}^2+a^2)^2+ma^2}\over{(r_{\infty}^2+a^2)^2-mr_{\infty}^2}}}
\end{equation}

Having the surface gravity on the outer horizon, one immediately
obtains the temperature of the black hole
\begin{eqnarray}\label{temp}
\nonumber T&=&{{\kappa}\over{2\pi}}\\
&=&{{(r_{+}-r_{-})\sqrt{(r_{\infty}^2+r_{+}r_{-})^2+r_{+}r_{-}(r_{+}+r_{-})^2}}\over{2\pi r_{+}(r_{+}+r_{-})(r_{\infty}^2-r_{-}^2)}}\times\sqrt{{r_{\infty}^2-r_{+}^2}\over{r_{\infty}^2-r_{-}^2}}
\end{eqnarray}

\subsection{Angular Velocity and Entropy}\label{ss42}

Different angular-velocity-like quantities have appeared in
Subsection \ref{ss31} and \ref{ss41}: $\Omega_H$, $\tilde{\Omega}_H$
and $\tilde{\omega}$. Which is the physical one? Notice that on the
outer horizon $\tilde{\omega}=\tilde{\Omega}_H$, so the problem to
determine angular velocity on the outer horizon is reduced to a
choice between $\Omega_H$ and $\tilde{\Omega}_H$. Their difference
originates from different definitions of time and angular
coordinates. As we have mentioned, the metric form (\ref{squa
metric3}) has better behaviors at spatial infinity, that is, a
vanishing angular momentum and a satisfactory normalization of the
temporal component. Hence we are sure now a meaningful angular
velocity on the event horizon is $\tilde{\Omega}_H$, given by
(\ref{g00}), taking the form
\begin{equation}\label{ang vel}
\tilde{\Omega}_H={{2}\over{r_{+}+r_{-}}}\times\sqrt{{r_{-}(r_{\infty}^2-r_{+}^2)}\over{r_{+}(r_{\infty}^2-r_{-}^2)}}\times{{r_{\infty}^2+r_{+}^2+2r_{+}r_{-}}\over{\sqrt{(r_{\infty}^2+r_{+}r_{-})^2+r_{+}r_{-}(r_{+}+r_{-})^2}}}
\end{equation}

It is not surprising that we should take into account asymptotic
properties of metric at infinity when computing thermodynamic
quantities on the event horizon. The key point is that all
thermodynamic quantities are ``seen from infinity''.

In \cite{CC0603} it has been found that the area formula of entropy
still survives in spite of a deformation of the horizon. Hereby we
calculate the entropy of Kerr-IM black hole according to the area
formula. Starting with (\ref{squa metric1}), (\ref{squa metric2}),
(\ref{edd metric}) or (\ref{squa metric3}), one easily obtains a
unique consequence
\begin{eqnarray}\label{entropy}
\nonumber A&=&2\pi^2r_{+}(r_{+}+r_{-})^2\times{{r_{\infty}^2-r_{-}^2}\over{r_{\infty}^2-r_{+}^2}}\\
S&=&{A\over4}={{\pi^2r_{+}(r_{+}+r_{-})^2}\over2}\times{{r_{\infty}^2-r_{-}^2}\over{r_{\infty}^2-r_{+}^2}}
\end{eqnarray}

\section{The First Law of Thermodynamics}\label{s5}
\subsection{Abbott-Deser Mass}\label{ss51}

In the previous section, we argue that when calculating
thermodynamic quantities, the asymptotic behavior of metric should
be taken into consideration. Especially, a suitable form of metric
to derive those quantities must asymptotically normalize the
temporal components and annihilate angular momenta. The metric form
(\ref{squa metric3}) has these properties, whereas a more convenient
form would be an analogous one in terms of $\rho$. It can be
obtained from (\ref{squa metric2}), after coordinate transformation
(\ref{psit}), or from (\ref{squa metric3}), after coordinate
transformation (\ref{rho})
\begin{equation}\label{squa metric4}
ds^2=-\tilde{g}_{00}d\tilde{t}^2+Ud\rho^2+R^2(\sigma_1^2+\sigma_2^2)+{{(r_{\infty}^2+a^2)^2+ma^2K^4}\over4(r_{\infty}^2+a^2)K^2}(\tilde{\sigma}_3-\tilde{\omega}d\tilde{t})^2
\end{equation}
in which $U$, $R$ and $K$ are given by (\ref{asym param}).
$\tilde{g}_{00}$ and $\tilde{\omega}$ are the same as those in
(\ref{g00}), also taking the form
\begin{eqnarray}\label{omiga}
\nonumber \tilde{g}_{00}&=&{{(r_{\infty}^2+a^2)^2-m(r_{\infty}^2+a^2)K^2+ma^2K^4}\over{(r_{\infty}^2+a^2)^2+ma^2K^4}}\times{{(r_{\infty}^2+a^2)^2+ma^2}\over{(r_{\infty}^2+a^2)^2-mr_{\infty}^2}}\\
\nonumber \tilde{\omega}&=&\left[{{2maK^4}\over{(r_{\infty}^2+a^2)^2+ma^2K^4}}-{{2ma}\over{(r_{\infty}^2+a^2)^2+ma^2}}\right]\times\sqrt{{{(r_{\infty}^2+a^2)^2+ma^2}\over{(r_{\infty}^2+a^2)^2-mr_{\infty}^2}}}\\
\end{eqnarray}

In Abbott and Deser's definition, conserved charges are associated
with isometries of the asymptotic geometry which is supposed to be
the vacuum of the system \cite{AD82}. In Kerr-IM space-time,
corresponding to (\ref{squa metric4}), the asymptotic geometry is
described by (\ref{asym metric2}).

Taking $\bar{\xi}^{\mu}\partial_{\mu}=\partial_{\tilde{t}}$ as the
canonically normalized time-like Killing vector, and repeating the
computation done in \cite{CC0603}, one is led to the generalized
Abbott-Deser mass of the Kerr-IM black hole
\begin{eqnarray}\label{kerr mass4}
\nonumber M_{Kerr-IM}&=&{\pi\over4}\times{{(r_{\infty}^2+a^2)^2-ma^2}\over{\sqrt{(r_{\infty}^2+a^2)^2+ma^2}}}\times{{(r_{\infty}^2+a^2)^2+m(r_{\infty}^2+2a^2)}\over{(r_{\infty}^2+a^2)\sqrt{(r_{\infty}^2+a^2)^2-mr_{\infty}^2}}}\\
\nonumber &=&{\pi\over4}\times{{(r_{\infty}^2+r_{+}r_{-})^2-r_{+}r_{-}(r_{+}+r_{-})^2}\over{\sqrt{(r_{\infty}^2+r_{+}r_{-})^2+r_{+}r_{-}(r_{+}+r_{-})^2}}}\\
&&\times{{(r_{\infty}^2+2r_{+}r_{-}+r_{+}^2)(r_{\infty}^2+2r_{+}r_{-}+r_{-}^2)}\over{(r_{\infty}^2+r_{+}r_{-})\sqrt{(r_{\infty}^2-r_{+}^2)(r_{\infty}^2-r_{-}^2)}}}
\end{eqnarray}
This mass and the mass of IM black hole obtained in \cite{CC0603}
will be analyzed in Subsection \ref{ss54}.

\subsection{Angular Momentum and the First Law}\label{ss52}

Starting with (\ref{squa metric4}), after a few calculations, the
Komar integrals give the angular momenta
\begin{eqnarray}\label{ang momen}
\nonumber J_{\phi}&=&0\\
J&=&J_{\tilde{\psi}}={\pi\over4}ma={\pi\over4}\sqrt{r_{+}r_{-}}(r_{+}+r_{-})^2
\end{eqnarray}

The first law of black hole thermodynamics in our case
\begin{equation}\label{first law}
dM-TdS-\tilde{\Omega}_HdJ=0
\end{equation}
involves five parameters. All of them have been obtained now. To
check the first law, one might view $r_{+}$, $r_{-}$ as variables
while keep $r_{\infty}$ as a constant. But that will break (\ref{first
law}). Remember that for Kerr-IM space-time the geometric parameter
at infinity is not $r_{\infty}$ but ${r'}_{\infty}$, as we have
explained in Subsection \ref{ss23}. In deed, if we fix
${r'}_{\infty}$ rather than $r_{\infty}$, the first law of black
hole thermodynamics (\ref{first law}) is obeyed.

\subsection{The Counter-term Method}\label{ss53}

For a space-time with an asymptotic boundary topology $R\times
S^1\hookrightarrow S^2$, a simple counter-term in gravitational
action is introduced by Mann and Stelea \cite{MS0511}. As a matter
of fact, asymptotic structures are the same for IM black holes and
Kaluza-Klein monopoles, the latter of which are treated in
\cite{MS0511}. An advantage of the counter-term method over the
background subtraction method is that it is independent of reference
backgrounds hence leads to a unique result. The stress tensor for
the asymptotically flat space-times was proposed for the first time
by Astefanesei and Radu in \cite{AR0509}. For different boundary
topologies there exist different counter-terms, but the form of the
stress tensor is generic.

We still start with (\ref{squa metric4}). After a substantial amount
of calculations, we find the non-vanishing components of stress
tensor
\begin{eqnarray}\label{count term}
\nonumber 8\pi GT^{\tilde{t}}_{~\tilde{t}}&=&{{(r_{\infty}^2+a^2)^4+m(r_{\infty}^2+a^2)^3-m^2a^2(r_{\infty}^2+2a^2)}\over{4[(r_{\infty}^2+a^2)^2+ma^2]\sqrt{(r_{\infty}^2+a^2)^3-mr_{\infty}^2(r_{\infty}^2+a^2)}}}{{1}\over{r^2}}+\mathcal{O}({{1}\over{r^3}})\\
\nonumber 8\pi GT^{\tilde{t}}_{~\phi}&=&-{{ma}\over{4}}\sqrt{{r_{\infty}^2+a^2}\over{(r_{\infty}^2+a^2)^2+ma^2}}{{\cos\theta}\over{r^2}}+\mathcal{O}({{1}\over{r^3}})\\
\nonumber 8\pi GT^{\tilde{t}}_{~\tilde{\psi}}&=&-{{ma}\over{4}}\sqrt{{r_{\infty}^2+a^2}\over{(r_{\infty}^2+a^2)^2+ma^2}}{1\over{r^2}}+\mathcal{O}({{1}\over{r^3}})\\
\nonumber 8\pi GT^{\theta}_{~\theta}&=&{{(r_{\infty}^2+a^2)^4-m(r_{\infty}^2-5a^2)(r_{\infty}^2+a^2)^2-m^2(r_{\infty}^4+3a^2r_{\infty}^2+a^4)}\over{32[(r_{\infty}^2+a^2)^3-mr_{\infty}^2(r_{\infty}^2+a^2)]}}{{1}\over{r^3}}+\mathcal{O}({{1}\over{r^4}})\\
\nonumber 8\pi GT^{\phi}_{~\phi}&=&{{(r_{\infty}^2+a^2)^4-m(r_{\infty}^2-5a^2)(r_{\infty}^2+a^2)^2-m^2(r_{\infty}^4+3a^2r_{\infty}^2+a^4)}\over{32[(r_{\infty}^2+a^2)^3-mr_{\infty}^2(r_{\infty}^2+a^2)]}}{{1}\over{r^3}}+\mathcal{O}({{1}\over{r^4}})\\
\nonumber 8\pi GT^{\tilde{\psi}}_{~\tilde{t}}&=&ma\left[{{r_{\infty}^2+a^2}\over{(r_{\infty}^2+a^2)^2+ma^2}}\right]^{3\over2}{{1}\over{r^2}}+\mathcal{O}({{1}\over{r^3}})\\
\nonumber 8\pi GT^{\tilde{\psi}}_{~\phi}&=&{{2(r_{\infty}^2+a^2)^4-m(r_{\infty}^2+a^2)^3+m^2a^2(r_{\infty}^2-a^2)}\over{4[(r_{\infty}^2+a^2)^2+ma^2]\sqrt{(r_{\infty}^2+a^2)^3-mr_{\infty}^2(r_{\infty}^2+a^2)}}}{{\cos\theta}\over{r^2}}+\mathcal{O}({{1}\over{r^3}})\\
8\pi GT^{\tilde{\psi}}_{~\tilde{\psi}}&=&{{2(r_{\infty}^2+a^2)^4-m(r_{\infty}^2+a^2)^3+m^2a^2(r_{\infty}^2-a^2)}\over{4[(r_{\infty}^2+a^2)^2+ma^2]\sqrt{(r_{\infty}^2+a^2)^3-mr_{\infty}^2(r_{\infty}^2+a^2)}}}{{1}\over{r^2}}+\mathcal{O}({{1}\over{r^3}})
\end{eqnarray}

Associated with the Killing vector $\partial_{\tilde{t}}$, the conserved
mass is the same as (\ref{kerr mass4}). Corresponding to
$\partial_{\phi}$ and $\partial_{\tilde{\psi}}$ respectively, the angular
momenta are given by (\ref{ang momen}).

\subsection{Comments and Conjectures}\label{ss54}
In a recent paper \cite{CC0603}, Cai et.al. have got an
expression similar to (\ref{kerr mass4}). In their notations, the IM
black hole mass is
\begin{equation}\label{im mass4}
M_{IM}=M_{CT}=M_{AD}={{\pi\left(r_{\infty}^4-3r_{+}^2r_{-}^2+r_{\infty}^2(r_{+}^2+r_{-}^2)\right)}\over{4\sqrt{(r_{\infty}^2-r_{+}^2)(r_{\infty}^2-r_{-}^2)}}}
\end{equation}
Besides a background subtraction method, i.e., the generalized
Abbott-Deser method \cite{CLP0510}, they also employed the
counter-term method \cite{MS0511}.

Both (\ref{kerr mass4}) and (\ref{im mass4}) grow as $r_{\infty}$
increases, that is, as the size of the compactified dimension
increases. This is a little counterintuitive. More surprisingly,
when the black hole disappears, i.e., in the ``empty'' limit
$r_{+}=r_{-}=0$, both masses tend to
\begin{equation}\label{mon mass4}
M_{mon}={{\pi r_{\infty}^2}\over{4}}
\end{equation}

We observe that (\ref{mon mass4}) is just the mass expression for
Kaluza-Klein monopoles in four dimensions \cite{Sorkin83,MS0511}.
Furthermore, if we set $r_{+}=0$ and $r_{-}=0$, the metrics for both
IM and Kerr-IM black holes simply reduce to the metric for
Kaluza-Klein monopoles \cite{Sorkin83}, whose geometry is perfectly
regular in five dimensions. This indicates the physical meaning of
(\ref{kerr mass4}) and (\ref{im mass4}): They express the masses of
the Kaluza-Klein black holes in the sight of four dimensions. The
Kaluza-Klein charges \cite{Sorkin83} account for their
counterintuitive behaviors. In a five-dimensional point of view,
there are no Kaluza-Klein charges, and (\ref{mon mass4}) is the
contribution of energy from background geometry. A further
observation is that if we subtract (\ref{mon mass4}) from (\ref{im
mass4}), the first law of thermodynamics checked in \cite{CC0603}
will not be affected, for they have fixed the value of $r_{\infty}$.

Strictly speaking, the expressions (\ref{kerr mass4}) and (\ref{im
mass4}) for mass are reliable only in the case of a finite
$r_{\infty}$ value, as we remarked in the end of Subsection
\ref{ss23}. Yet we want to extrapolate the result to the limit
$r\rightarrow r_{\infty}$. Then we propose the following conjectures
for an arbitrary value of positive $r_{\infty}$.

\begin{enumerate}
\item The formula (\ref{im mass4}) represents the mass of IM black
holes observed in four dimensions. Subtracting the monopole
contribution (\ref{mon mass4}), we conjecture the mass well-defined
in five dimensions
\begin{equation}\label{im mass5}
\mathcal{M}_{IM}={{\pi\left(r_{\infty}^4-3r_{+}^2r_{-}^2+r_{\infty}^2(r_{+}^2+r_{-}^2)\right)}\over{4\sqrt{(r_{\infty}^2-r_{+}^2)(r_{\infty}^2-r_{-}^2)}}}-{{\pi
r_{\infty}^2}\over{4}}
\end{equation}
\item Similarly, the formula (\ref{kerr mass4}) represents the mass of Kerr-IM black
holes observed in four dimensions. Subtracting the monopole
contribution (\ref{mon mass4}), we conjecture the mass well-defined
in five dimensions
\begin{eqnarray}\label{kerr mass5}
\nonumber \mathcal{M}_{Kerr-IM}&=&{\pi\over4}\times\left[{{(r_{\infty}^2+a^2)^2-ma^2}\over{\sqrt{(r_{\infty}^2+a^2)^2+ma^2}}}\times{{(r_{\infty}^2+a^2)^2+m(r_{\infty}^2+2a^2)}\over{(r_{\infty}^2+a^2)\sqrt{(r_{\infty}^2+a^2)^2-mr_{\infty}^2}}}\right.\\
&&\left.-{{(r_{\infty}^2+a^2)^2+ma^2}\over{r_{\infty}^2+a^2}}\right]
\end{eqnarray}
Here we have considered that the size of the compactified dimension
is controlled by ${r'}_{\infty}$ instead of $r_{\infty}$.
\end{enumerate}

Obviously, the masses $\mathcal{M}_{IM}$ and $\mathcal{M}_{Kerr-IM}$
also satisfy the first law of balck hole thermodynamics in five
dimensions. To inspect and improve our conjectures, one may check
the law in four dimensions with $M_{IM}$ and $M_{Kerr-IM}$, by
including monopole charges and allowing ${r'}_{\infty}$ variable. It
is remarkable that in the ``flat'' limit $r\rightarrow r_{\infty}$,
we find $\mathcal{M}_{IM}\rightarrow{{3\pi(r_{+}^2+r_{-}^2)}\over8}$
and $\mathcal{M}_{Kerr-IM}\rightarrow{{3\pi m}\over8}$ respectively,
which are the masses of ordinary five-dimensional
Reissner-Nordstr\"om black holes and Kerr black holes respectively.
In the ``empty'' limit $r_{+}=r_{-}=0$, both $\mathcal{M}_{IM}$ and
$\mathcal{M}_{Kerr-IM}$ vanish. It would be interesting and
important to look for a new background and a new counter-term to
``fundamentally'' produce the masses (\ref{im mass5}) and (\ref{kerr
mass5}), and at the same time the angular momenta (\ref{ang momen}).
But it is not the aim of this article.

\section{Summary}\label{s6}
In this paper, we have extended the recently found static
Kaluza-Klein black holes with squashed horizons \cite{IM0510} to
rotating counterparts. Our investigations are restricted to black
holes with two equal angular momenta. These counterparts are
geodesic complete and free of naked singularities. The thermodynamic
quantities have been calculated and the first law has been checked.
The angular momenta are computed using Komar integrals, then
reproduced with the counter-term method. We calculated the mass by
both background subtraction and counter-term method, and found a
counterintuitive behavior as the compactified dimension expands.
This behavior is also hidden in the mass of static black holes. We
attributed it to Kaluza-Klein charges and gave some comments on it.

{\bf Acknowledgement}: We would like to thank Li-Ming Cao, Miao Li,
Jian-Xin Lu, Wei Song and Yushu Song for useful discussions and kind
help. We are especially grateful to Rong-Gen Cai for a careful
reading of the manuscript and many helpful discussions.

\vfill\eject

\end{document}